# Singlet-triplet spin blockade and charge sensing in a few-electron double quantum dot

A. C. Johnson, J. R. Petta, C. M. Marcus
*Department of Physics, Harvard University, Cambridge, Massachusetts 02138*

M. P. Hanson, A. C. Gossard
*Department of Materials, University of California, Santa Barbara, California 93106*

Singlet-triplet spin blockade in a few-electron lateral double quantum dot is investigated using simultaneous transport and charge-sensing measurements. Transport from the (1,1) to the (0,2) electron occupancy states is strongly suppressed relative to the opposite bias [(0,2) to (1,1)]. At large bias, spin blockade ceases as the (0,2) triplet state enters the transport window, giving a direct measure of exchange splitting as a function of magnetic field. A simple model for current and steady-state charge distribution in spin-blockade conditions is developed and found to be in excellent agreement with experiment.

Great progress has been made in engineering solid-state systems that exhibit quantum effects, providing new tools for probing fundamental problems in many-body physics as well as new device technologies. In semiconductor quantum dots, small numbers of confined electrons can be manipulated using electrostatic gates with surprising ease [1, 2]. For the case of two electrons in the dot (quantum dot "helium"), Pauli exclusion and exchange induce a splitting between the spin singlet and triplet states that can be controlled by gates and magnetic fields [3]. In double dots, a consequence of this splitting is current rectification, in which transitions from the (1,1) to the (0,2) state (ordered pairs indicate electron occupancy in each dot) is blockaded, while the opposite bias case, involving transitions from (0,2) to (1,1) proceeds freely. Rectification is a direct consequence of spin selection rules [4].

Spin blockade of this type can be understood by considering positive and negative bias transport in a double dot containing one electron in the right dot, as indicated in Figs. 1(b,c). An electron of any spin can enter the left dot, making either a (1,1) singlet or triplet, these states being nearly degenerate for weak interdot tunneling [5]. In contrast, the right dot can accept an electron only to make a (0,2) singlet. At positive bias (Fig. 1b) current can flow: an electron enters the right dot to make a (0,2) singlet, tunnels to the (1,1) singlet, and escapes. At negative bias (Fig. 1c), an electron can enter the left dot and form a (1,1) triplet state. A transition from the (1,1) triplet to the (0,2) singlet is forbidden by conservation of spin and transport is blocked.

In this Letter we report a detailed investigation of spin blockade through a lateral few-electron double-dot system, measured using both transport and charge sensing by a nearby quantum point contact (QPC) to detect the charge arrangement during blockade [6]. We observe that transmission through double-dot states containing two electrons is strongly rectified, while transmission of the first electron is symmetric in bias. Negative-bias blockade is truncated when the (0,2) triplet state enters the bias window, allowing the magnetic field dependence of the singlet-triplet splitting to be measured from both transport and charge sensing. Simple rate-equation models for transport and charge distributions reproduce key features in both types of data, and allow relative tunnel rates to be extracted.

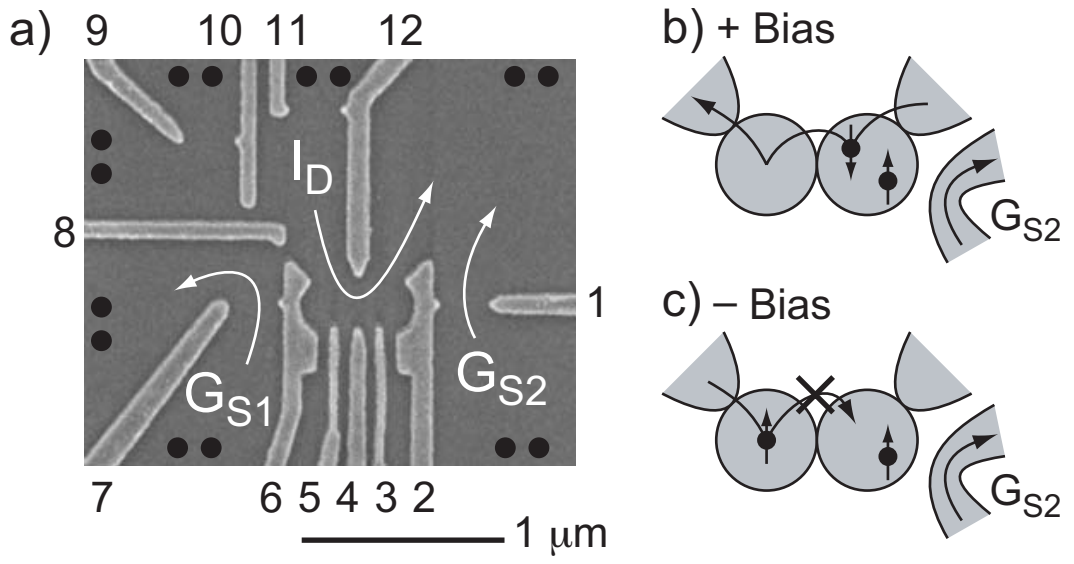

FIG. 1. (a) Electron micrograph of a device identical in design to the one measured. Gates 2-6 and 12 define the double dot, 1 and 7 form QPC charge sensors, 8 separates the left QPC and double dot current paths, and 9-11 are unused. Black spots (●) denote ohmic contacts. The origin of spin blockade, when the left dot has 0-1 electrons and the right dot 1-2 electrons, is illustrated in (b) and (c). Opposite spins represent a singlet, same spins a triplet. The left dot accepts any spin but the right can only form a spin singlet, blocking negative bias current once the wrong spin occupies the left dot. A charge sensor ($G_{S2}$) also registers the blockade, as a second electron is sometimes in the right dot at positive bias, but not at negative bias.

Spin blockade of transport, arising from a variety of mechanisms, has been investigated previously in quantum dot systems [2, 7]; the mechanism responsible for the present spin blockade was investigated in vertical structures in Ref. [4]. The lateral, gate-defined structure we investigate has some advantages over vertical structures by allowing independent tuning of all tunnel barriers so that sequential tunneling with arbitrary dot occupations can be explored. QPC charge sensors provide additional information, including the average charge distribution when transport is absent in the spin blockade regime.

The sample, shown in Fig. 1(a), is fabricated on a $GaAs/Al_{0.3}Ga_{0.7}As$ heterostructure with two-dimensional electron gas (density $2\times10^{11}$ cm$^{-2}$, mobility $2\times10^{5}$ cm$^2$/Vs) 100 nm below the wafer surface, patterned with Ti/Au top gates. Gates 2-6 and 12 define a double quantum dot in which each dot can be tuned from zero to several electrons

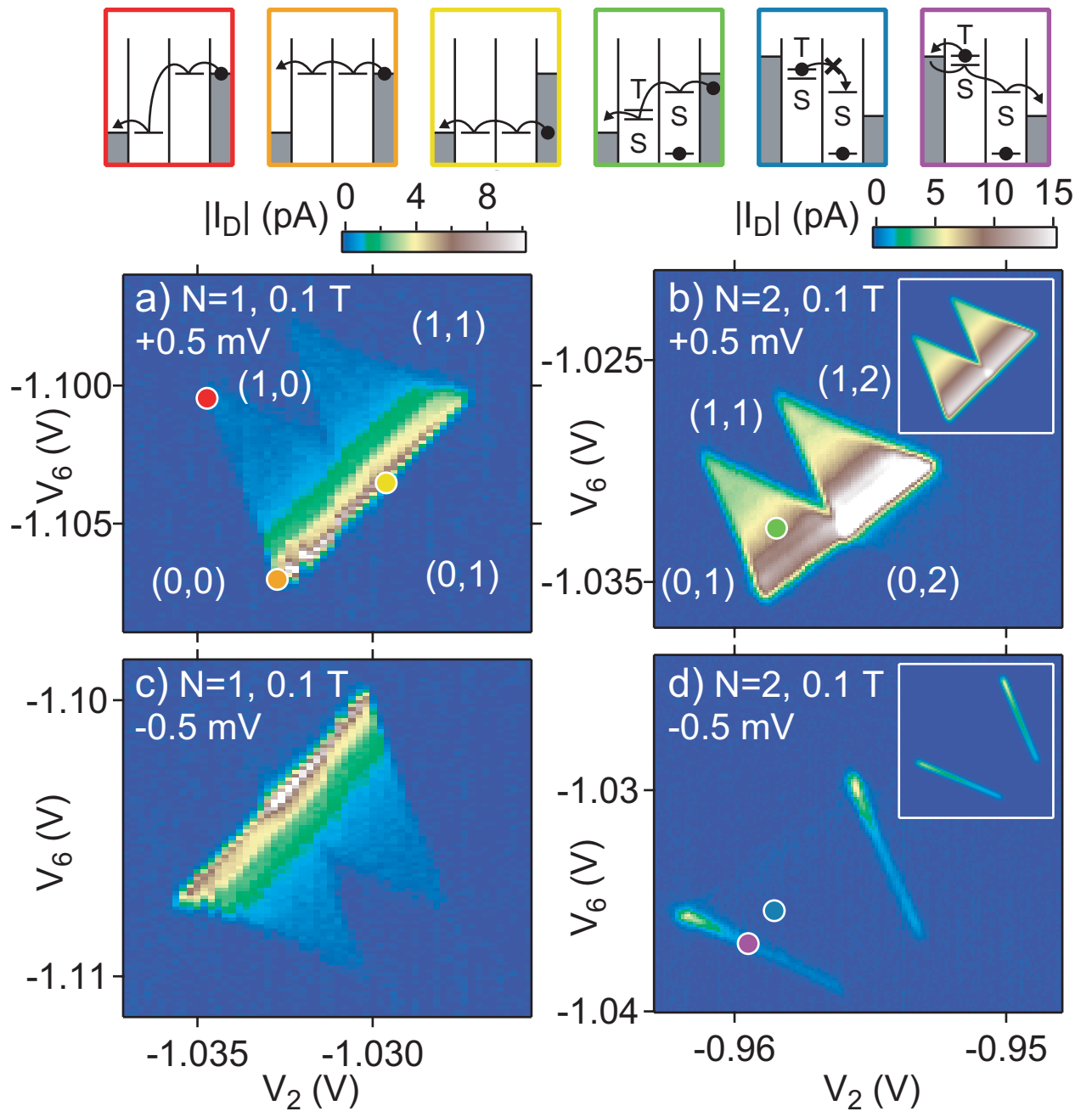

FIG. 2. (a) Magnitude of current $I_D$ as a function of $V_2$ and $V_6$ across the (1,0) to (0,1) transition at 0.5 mV bias. Ordered pairs ($m$,$n$) denote electrons on the left ($m$) and right ($n$) dots, with $N$ total electrons present during interdot tunneling. The red, orange, and yellow diagrams illustrate the level alignments bounding a bias triangle. The same configuration at -0.5 mV bias, (c), shows almost perfect symmetry. (b) and (d) show the equivalent data at the (1,1) to (0,2) transition. Current flows freely at positive bias, as depicted in the green diagram. Negative bias current is suppressed by spin blockade (blue diagram) except on the lower (purple diagram) and upper edges. Insets to (b) and (d) show results of a rate equation model which captures most features of the data (see text).

[8]. Gates 1 and 7 define QPCs whose conductance is most sensitive to the charge on the right and left dots respectively. Gate 8 isolates the current path of the double dot from the left QPC, while the double dot and right QPC share a common ground. Gates 9-11 are not energized. Current through the double dot ($I_D$) is measured in response to a dc voltage on the left reservoir. A small ac excitation (6 μV at 27 Hz) allows lock-in measurement of differential conductance. Conductances of QPC charge sensors ($G_{S1,2}$) are measured simultaneously with separate lock-in amplifiers (1 nA current bias at 93 and 207 Hz). Base electron temperature is $T_e$~135 mK, measured from Coulomb blockade diamonds. Two devices were measured and showed qualitatively similar behavior; data from one device are presented.

Figure 2 shows $I_D$ at ±0.5 mV bias as a function of gate voltages $V_2$ and $V_6$, which primarily control the energy levels in the right and left dots. Figures 2(a,c) were measured near the conductance resonance of the first electron, with ($m$,$n$) indicating the charge states surrounding the resonance. At positive bias, finite current is measured within two overlapping triangles in gate voltage space, satisfying the inequalities $\mu_R \geq \varepsilon_R \geq \varepsilon_L \geq \mu_L$ or $\mu_R \geq \varepsilon_R + E_m \geq \varepsilon_L + E_m \geq \mu_L$. Here $\mu_{L,R}$ are the chemical potentials of the leads, $\varepsilon_{L,R}$ are the energies to add an electron to the ground state of either dot, and the mutual charging energy $E_m$ is the extra energy to add an electron to one dot with the other dot occupied [9]. The first set of inequalities defines the lower, or *electron triangle*, where, starting at (0,0), an electron hops through the dots from one lead to the other. The second inequalities define the upper, or *hole triangle*, where, starting at (1,1), a hole hops across the dots. Electron and hole processes involve the same three tunneling events, only their order changes. Schematics at the top of Fig. 2 depict the energy level alignments at the vertices of the electron triangle in Fig. 2(a). Within the triangles, current depends primarily on the detuning $\Delta = \varepsilon_L - \varepsilon_R$ of one-electron states, with a maximum current at $\Delta=0$, demonstrating that interdot tunneling is strongest at low energy loss [10]. At negative bias (Fig. 2c) the triangles flip and current changes sign, but otherwise these data mimic the positive bias case (Fig. 2a).

The corresponding data with another electron added to the right dot is shown in Figs. 2 (b,d). At positive bias, the data qualitatively resemble the one-electron case. However, at negative bias the current is nearly zero, except along the outermost edges of the electron and hole triangles. Referring to the diagrams above Fig. 2, at positive bias, current proceeds freely from right to left through singlet states (green diagram). At negative bias, an electron enters the left dot into either the (1,1) singlet or triplet. If it enters the (1,1) singlet, it may continue through the (0,2) singlet. However, once an electron enters the (1,1) triplet, it can neither continue to the right (into the (0,2) singlet) nor go back into the left lead because it is below the Fermi level and the hole it left quickly diffuses away. Thereafter, negative-bias transport requires a spin flip or a second-order spin exchange process with one of the leads. Insofar as these processes are relatively slow, transport in this direction is blockaded.

Along the outer edge of the lower (electron) triangle, where transport is observed in the negative-bias direction (Fig. 2(d), purple dot), an electron trapped in the (1,1) triplet state is within the thermal window of the left lead and will occasionally exchange with another electron possibly loading the (1,1) singlet, which can the move to the right, through the (0,2) singlet, and contribute to current. An analogous mechanism in the hole channel allows negative-bias current along the upper edge of the hole triangle: with transitions from (1,1) to (1,2) within the thermal window, the blockade created by an occupied (1,1) triplet can be lifted by adding an electron, making a (1,2) state, then removing it, possibly leaving a (1,1) singlet that can contribute to current.

A simple rate-equation model allows the spin-blockade picture to be quantitatively checked against transport data, and also indicates where charge resides in the double dot, which can be compared to charge sensing data. The model takes two degenerate levels in the left dot, representing the (1,1) singlet and triplet states, coupled equally to a thermally broadened left reservoir (i.e., ignoring the extra degeneracy of the triplet) and a single level of the right dot, representing the (0,2) singlet (assuming the (0,2) triplet is energetically inaccessible) coupled to the right reservoir.

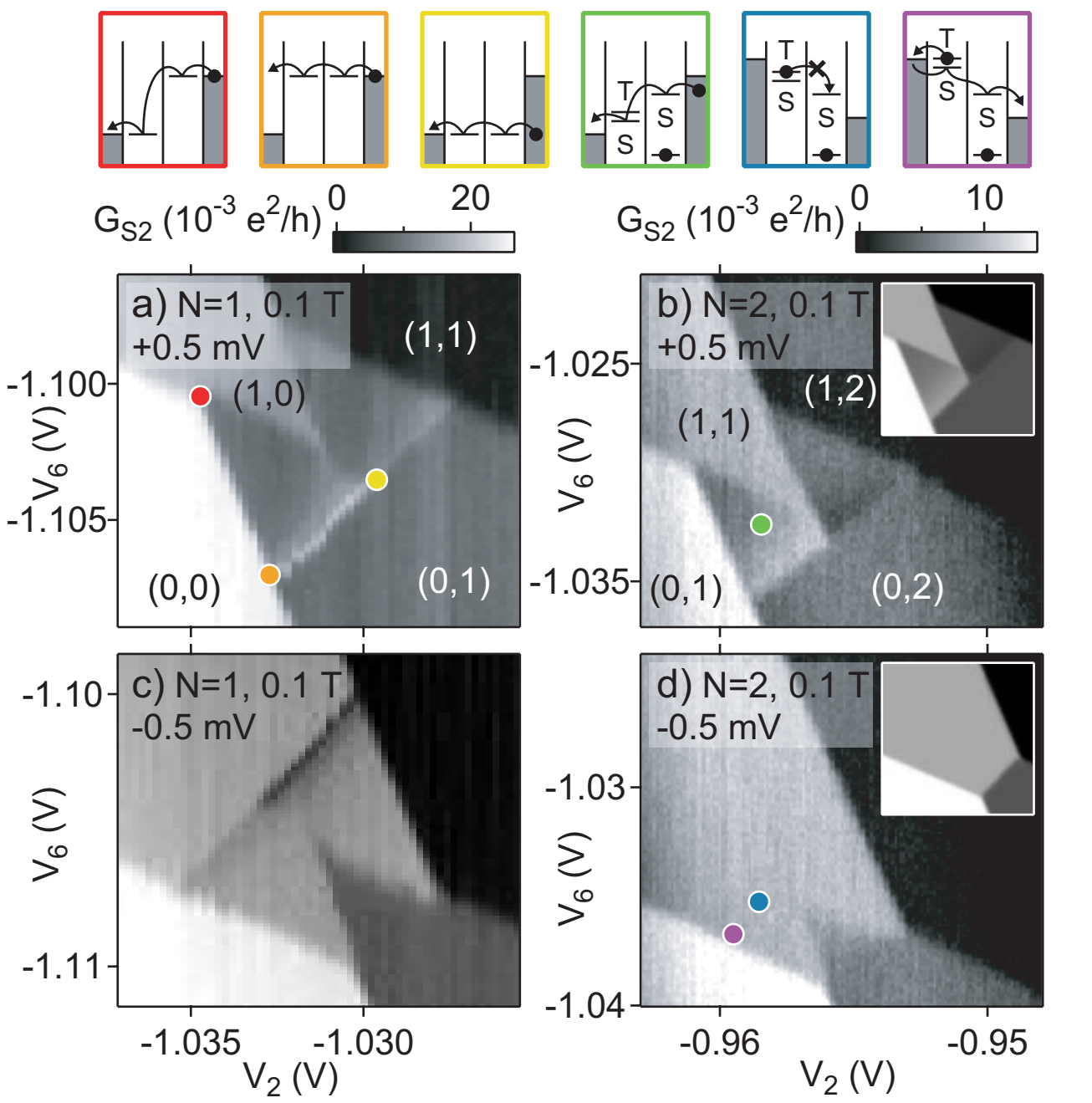

FIG. 3. Charge sensor signal $G_{S2}$ measured simultaneously with each panel of Fig. 2. A plane is subtracted from each panel to remove direct gate-QPC coupling. The first electron, (a) and (c), again shows bias symmetry while the second, (b) and (d), is missing features at negative bias. The model used in Fig. 2 also reproduces the charge sensor data, as shown in the insets to (b) and (d), with slight disagreement due to second-order processes (see text).

The singlet levels are coupled by thermally activated inelastic tunneling, with the shape of the $\Delta$=0 peak inserted to match the positive-bias current data. Temperature, mutual charging energy, and the gate capacitances are determined from measurements. Calculated current is shown in the insets to Figs. 2(b) and (d). The model resembles the experimental data, with two minor exceptions: At positive bias, measured current is higher in the hole triangle than the electron triangle, implying that the dot-lead tunnel barriers are, in this case, more transparent with the other dot occupied. Also, the small but finite blockade current is absent in the model, as expected since the model contains only first-order, spin-conserving processes.

Figure 3 shows the charge sensor data $G_{S2}$ versus $V_2$ and $V_6$, acquired simultaneously with each panel in Fig. 2. A plane is subtracted from each data set to remove direct coupling between the gates and the QPC, leaving only the effect of the average dot occupations. Away from the bias triangles we see plateaus for each stable charge state, which are used to calibrate the response. In Figs. 3(a) and (c), QPC conductance jumps $\Delta G_R$ = 0.016 $e^2/h$ due to a charge in the right dot, and $\Delta G_L$ = 0.008 $e^2/h$ due to the left. These values vary between data sets, but this QPC is always about twice as sensitive to the closer dot.

Within each bias triangle, the sensing signal varies with the fraction of time an electron spends in each charge state. Consider Fig. 3(a), the one-electron positive bias data. As with transport through the dot, charge sensing is primarily dependent on detuning, $\Delta$. For small interdot tunneling, the system rests mainly in (0,1), thus at large detuning the sensing signal matches the (0,1) plateau. In the electron triangle, as detuning decreases and interdot tunneling increases, the system spends more time in (1,0) and (0,0). Both increase the right QPC sensor conductance. In the hole triangle, the system accesses (1,0) and (1,1), which respectively increase and decrease the right sensing signal.

Assuming the same tunnel rates to the leads in the electron and hole triangles, lead asymmetry can be quantified by comparing the rate-equation model to the sensor signal in the two triangles. Occupancies of participating charge states are found in terms of tunnel rates (determined from matching the model to transport data), and then calculating the QPC response in each triangle. The ratio of right lead to left lead tunnel rates is given by $\Gamma_R / \Gamma_L$ = $(S\text{-}1\text{-}S\alpha)/(1+(S\text{-}1)\alpha)$, where $\alpha$ is the ratio of the $\Delta$=0 sensor peak height in the hole triangle to that in the electron triangle and $S$ is the QPC sensitivity ratio $\Delta G_R / \Delta G_L$. From Fig. 3(a), $\alpha$ = 0.35±0.05, giving $\Gamma_R / \Gamma_L$ = 0.23±0.08. Similar analysis at negative bias (Fig. 3c) yields $\Gamma_R / \Gamma_L$ = 0.35±0.10. In the positive bias two-electron case (Fig. 3b), the hole triangle shows a negative sensor change, indicating that the left barrier is more opaque than the right. It is not possible to further quantify this ratio, as we know from Fig. 2(b) that tunnel rates in the two triangles differ.

In the spin blockade region (Fig. 3d, near the blue dot), no sensor variation is seen, confirming that the system is trapped in (1,1). By including lead asymmetry and QPC sensitivities, the rate-equation model used for the insets to Figs. 2 (b) and (d) yields charge sensor signals as well. These are shown in the insets to Figs. 3(b) and (d), and again the agreement is good. The one discrepancy is between the two triangles at negative bias. The model shows a thermally broadened transition from (1,1) to (0,2) at a detuning equal to the bias. This is equivalent to strictly zero interdot tunneling, in which case the system has a ground state and occupations mimic a zero-bias stability diagram. However, between the bias triangles the only mechanism for equilibration is a second order process of each dot exchanging an electron with its lead. This is slow enough to occur on par with the second order and spin-flip processes noted above which circumvent the blockade, so the sensor shows a mixture of (1,1) and (0,2).

Figure 4 illustrates the features that arise at dc bias larger than $\Delta_{ST}$, the (0,2) singlet-triplet energy splitting. Panels (a)-(d) show $I_D$ and $G_{S2}$ versus voltages $V_2$ and $V_6$ at ±1 mV bias and a perpendicular magnetic field $B_\perp$=0.9 T. At negative bias, spin blockade is lifted when the (1,1) triplet is raised above the (0,2) triplet (pink diagram), so in Figs. 4(c) and (d), current turns on and steady-state populations shift where $\Delta > \Delta_{ST}$. At positive bias, current increases (Fig. 4a) and populations shift (Fig. 4b) when the Fermi level in the right lead accesses the (0,2) triplet (black diagram). Thus, $\Delta_{ST}$ can be measured using either current or charge sensing and either sign of bias, with the energy scale calibrated by equating the triangle size to the dc bias [9]. Current and sensing give consistent values of $\Delta_{ST}$, but different splittings are measured at different biases and gate voltages, presumably reflecting real changes in $\Delta_{ST}$ as these parameters are tuned. The two measurements in Fig. 4(a) give 480 and 660 μeV. The negative bias measurement gives ~520±50 μeV in Fig. 4(c). The fact that positive bias measurements differ more than negative bias measurements implies that occupation of the left dot has a strong effect on the right dot levels.

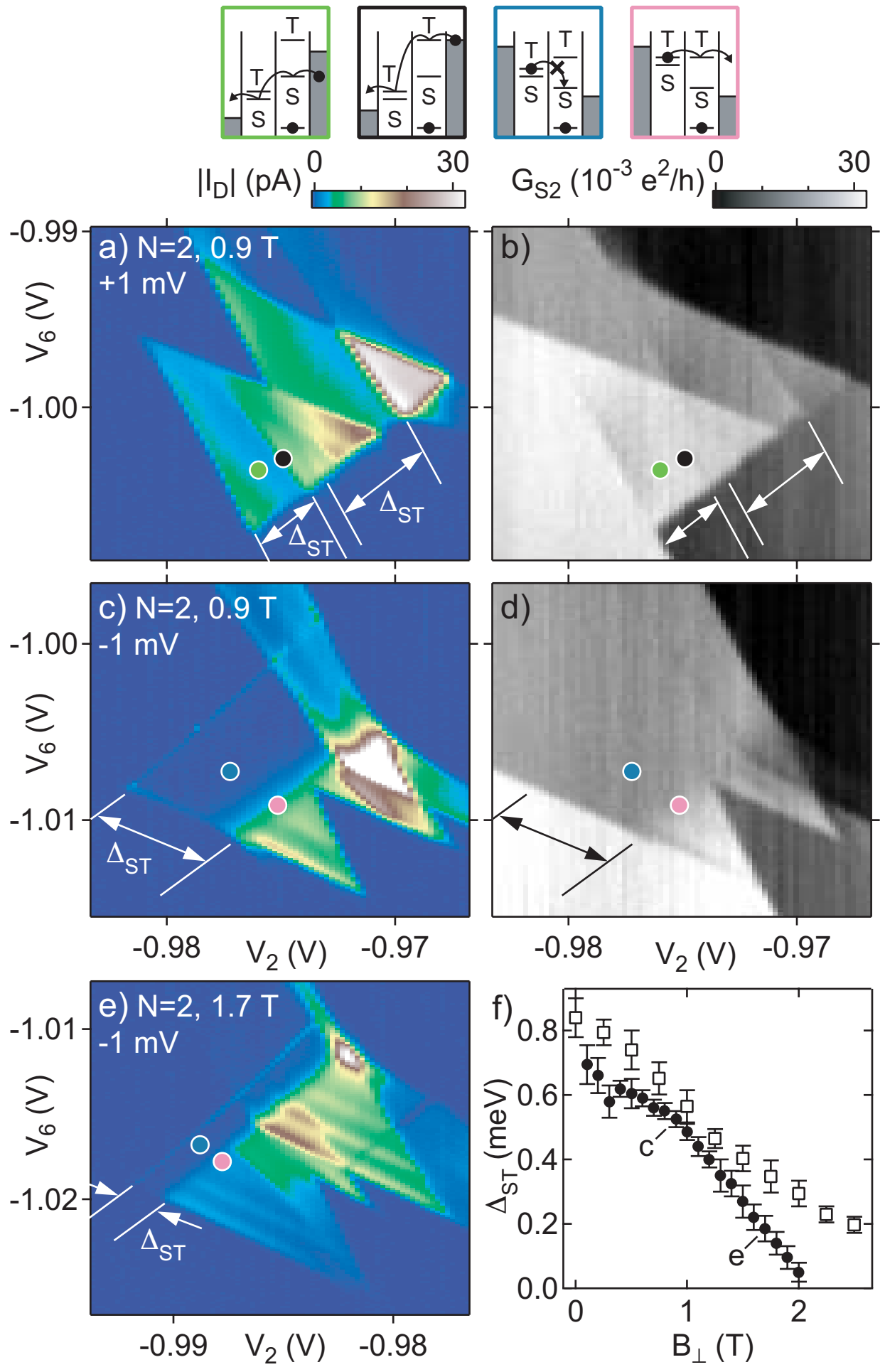

FIG. 4. Measurements of the singlet-triplet splitting, $\Delta_{ST}$. At +1 mV bias, $I_D$ (a) and $G_{S2}$ (b) as a function of $V_2$ and $V_6$ at the (1,1) to (0,2) transition show new features where triplet current is allowed (black diagram and spot) rather than just singlet current (green). At negative bias, $I_D$ (c) turns on and $G_{S2}$ (d) changes when triplet states break the spin blockade (pink vs. blue). Several $\Delta_{ST}$ values are measured from (a)-(d), which we attribute to gate voltage dependence of $\Delta_{ST}$. As $B_\perp$ increases, (e), $\Delta_{ST}$ decreases rapidly. The dependence of $\Delta_{ST}$ on $B_\perp$ is shown in (f), with the solid and open series both taken at negative bias from different gate voltage configurations.

Increasing $B_\perp$ reduces $\Delta_{ST}$, bringing the negative bias current threshold closer to the Δ=0 edge. Figure 4(e) shows negative bias current at $B_\perp$=1.7 T, where a dramatic decrease in $\Delta_{ST}$ is seen compared to Fig. 4(c). Figure 4(f) shows $\Delta_{ST}$ as a function of $B_\perp$ based on negative-bias data at different gate voltage settings. For the open squares, no voltages but the swept $V_2$ and $V_6$ were changed during the field sweep. The tunnel barriers closed with increasing field, obscuring the measurement above 2.5 T, before the splitting reached zero. The sweep yielding the filled circles and panels (a)-(e), started from different gate voltages and gave $V_4$ and $V_{12}$ corrections quadratic in field to keep the barriers roughly constant. $\Delta_{ST}$ dropped until at 2.1 T no splitting was observed. A split feature was hinted at in the 2.5 T positive bias data, implying that the triplet had become the ground state, but again the signal vanished as field increased. Zeeman splitting, though it would be a small effect regardless, is entirely absent at negative bias as interdot transitions connect states with equal Zeeman energy.

We acknowledge useful discussions with Jacob Taylor and Amir Yacoby. This work was supported by the ARO under Grant Nos. DAAD55-98-1-0270 and DAAD19-02-1-0070, the DARPA QuIST program, the NSF under Grant No. DMR-0072777 and the Harvard NSEC, and iQuest at UCSB.